\documentclass[pdflatex,sn-basic]{sn-jnl}% Basic Springer Nature Reference Style/Chemistry Reference Style

\usepackage{algorithm}%
\usepackage{algorithmicx}%
\usepackage{algpseudocode}%
\usepackage{amsmath,amssymb,amsfonts}%
\usepackage{amsthm}%
\usepackage[title]{appendix}%
\usepackage{bm}
\usepackage{booktabs}%
\usepackage{graphicx}%
\usepackage{listings}%
\usepackage{manyfoot}%
\usepackage{mathrsfs}%
\usepackage{multirow}%
\usepackage{textcomp}%
\usepackage{xcolor}%
\theoremstyle{thmstyleone}%
\theoremstyle{thmstyletwo}%

\theoremstyle{thmstylethree}%

\begin{document}

\title[PNN for ionosphere estimation]{Exploring the usage of Probabilistic Neural Networks for Ionospheric electron density estimation}

\author*[1]{\fnm{Miquel} \sur{Garcia-Fernandez}}\email{miquel.garcia.fernandez@upc.edu}

\affil*[1]{\orgname{UPC/IonSat}, \orgaddress{\city{Barcelona}, \postcode{08034}, \state{Catalunya}, \country{Spain}}}

\abstract{
A fundamental limitation of traditional Neural Networks (NN) in predictive modelling
is their inability to quantify uncertainty in their outputs. In critical applications
like positioning systems, understanding the reliability of predictions is critical for
constructing confidence intervals, early warning systems, and effectively propagating
results. For instance, Precise Point Positioning in
satellite navigation heavily relies on accurate error models for ancillary data (orbits,
clocks, ionosphere, and troposphere) to compute precise error estimates. In addition,
these uncertainty estimates are needed to establish robust protection levels in safety
critical applications.

To address this challenge, the main objectives of this paper aims at exploring a
potential framework capable of providing both point estimates and associated uncertainty
measures of ionospheric Vertical Total Electron Content (VTEC). In this context, Probabilistic
Neural Networks (PNNs) offer a promising approach to achieve this goal. However,
constructing an effective PNN requires meticulous design of hidden and output layers,
as well as careful definition of prior and posterior probability distributions for network
weights and biases.

A key finding of this study is that the uncertainty provided by
the PNN model in VTEC estimates may be systematically underestimated. In
low-latitude areas, the actual error was observed to be as much
as twice the model's estimate. This underestimation is expected to be more
pronounced during solar maximum, correlating with increased VTEC values.

}

\keywords{
PNN, Neural Network, Bayesian Neural Network, Ionosphere, VTEC
}

\maketitle

\section{Introduction}\label{sec::introduction}

A fundamental limitation of traditional Neural Networks (NN) in predictive modelling
is their inability to quantify uncertainty in their outputs. In critical applications
like positioning systems, understanding the reliability of predictions is paramount
for constructing confidence intervals, early warning systems, and effectively
propagating results. For instance, Precise Point Positioning (PPP, see
\cite{zumberge1997precise}) in satellite
navigation heavily relies on accurate error models for ancillary data
(orbits, clocks, ionosphere, and troposphere) to compute precise error estimates
and establish robust protection levels. As an example, one of the main objectives
of the Galileo High Accuracy Service (HAS) Service Level 2 will be to provide the
necessary regional atmospheric delay corrections (and associated uncertainty) in
order to improve user positioning based on PPP strategies, most notably the
convergence time of the solution (see for instance \cite{juan2025iono4has}).

To address this challenge, the main objectives of this paper aims at
exploring a potential framework capable of providing both point estimates and associated
uncertainty measures of ionospheric Vertical Total Electron Content (VTEC).
Probabilistic Neural Networks (PNNs) offer a promising
approach to achieve this goal. However, constructing an effective PNN requires
meticulous design of hidden and output layers, as well as careful definition of
prior and posterior probability distributions for network weights and biases.

This introduction provides a review in terms of state-of-the-art in
PNN as well as the application of NN in ionospheric estimation of VTEC.

\subsection{Probabilistic Neural Network}

The weights and the biases in a NN are referred to as the network parameters
and are fit using a given input dataset and a certain criteria to maximize
likelihood (driven by a loss function) in a back-propagation strategy.
This type of neural network is referred to as \emph{point estimate neural network},
because once the network has undergone the fitting process, the parameters are
fixed. However, the outputs generated by such network do not provide any means
of uncertainty estimation and, additionally, the parameter estimation process
can also depend on the initialization process of the parameters
(e.g. values drawn from random variables in libraries such as TensorFlow
\cite{braiek2019tfcheck}) which will lead to different parameter values even for
the same training dataset.

\begin{figure}[h]
\centering
\includegraphics[width=0.7\columnwidth]{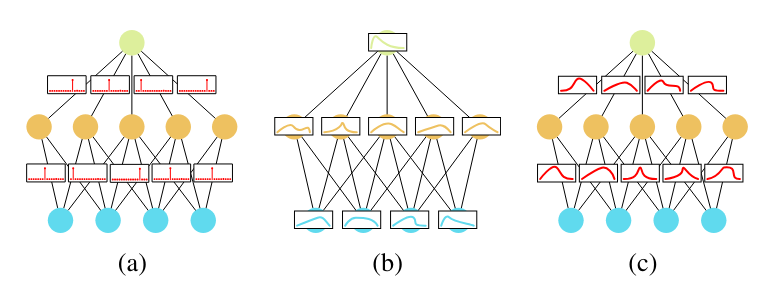}
\caption{Different types of probabilistic neural network concepts: (a) point
estimate neural network (i.e. non-probabilistic neural network), (b) neuron
activation driven through a stochastic distribution of probability and (c)
neuron weights and biases driven through a stochastic distribution of probability.
Source: Figure 3 of \cite{jospin2022hands}}\label{fig::jospin_pnn}
\end{figure}

In order to provide a solution that is able to address the shortcomings of the
traditional NN (fundamentally the lack of uncertainty provision to the model
estimates), probabilistic (or Bayesian) neural networks (PNN) have been proposed
(see for instance \cite{specht1990probabilistic}, \cite{mohebali2020probabilistic}
and, in particular, \cite{jospin2022hands} for a thorough description of PNNs).
The fundamental idea behind PNN is the introduction of stochastic components in
the model definition, either at the activation stage or at the parameters
(see Figure \ref{fig::jospin_pnn}). This means that these different components will have a stochastic behaviour:
in the case of the layer activation, the neurons will be activated or not
depending on a certain probability given by a Probability distribution.
Similarly, in the case of a PNN whose parameters are stochastics, the actual
values of the weights and biases will be drawn from a certain probability
distribution. The realisation of the stochastic elements is performed at every
prediction stage, this implies that different values of the output layer (i.e.
model estimates) will be generated at every prediction step.

The stochastic nature of the PNN allows the model user to obtain an estimate on
the uncertainty (see for instance \cite{gawlikowski2023survey} and
\cite{stahl2020evaluation}). Uncertainty in neural networks is usually divided
in two types (\cite{hullermeier2021aleatoric}): epistemic (i.e. systematic) and
aleatoric (i.e. statistical). The epistemic uncertainty is generated due to the
training process with incomplete datasets. An example of epistemic uncertainty,
in the context of ionosphere, would be the uncertainty of VTEC estimates during
maximum solar cycle when only data from minimum solar cycle has been considered
for the training stage. The epistemic uncertainty can be reduced by incorporating
new datasets for training (in the previous example, VTEC data from maximum solar
cycle periods). The aleatoric uncertainty comes from the randomness inherent in
the input data (i.e. noise), and cannot be in principle reduced by adding more
data to the model.

In more practical terms, in order to compute estimated uncertainties in neural
network models, \emph{ensembles} are used. These ensembles can be understood as
a collection of multiple models trained on the same dataset, but with different
initializations, parameters, or architectures. The model estimates are then built
by means of averaging these ensembles while the uncertainty can be obtained by
computing the dispersion of the different model estimates of the ensemble against
this average. In this context, several strategies can be followed to achieve
these ensembles:

\begin{itemize}
    \item Using a standard Neural Networks, ensembles can be achieved by means
    of “Bootstrap-aggregation” (see for instance \cite{lee2020bootstrap}, illustrated in Figure \ref{fig::bootstrap}),
    which in fact does not strictly require the usage of Probabilistic Neural Network,
    because in this case the original dataset is resampled $K$ times in order
    to obtain $K$ different model fits (each model with its own set of parameter ($\theta$)
    estimates. Each of these models will make a different prediction that can
    be combined in order to obtain a mean value and a dispersion metric
    (usually standard deviation) as a means to evaluate the uncertainty.

    \begin{figure}[h!]
    \centering
    \includegraphics[width=0.7\columnwidth]{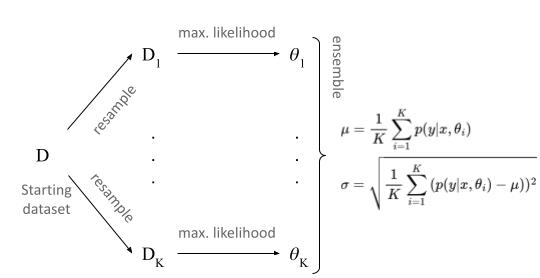}
    \caption{Bootstrap aggregation resample a dataset to obtain various models}\label{fig::bootstrap}
    \end{figure}

    \item A similar effect can be achieved with Bayesian or Probabilistic Neural Networks
    (\cite{jospin2022hands}, being the main difference the fact that the parameters
    are not estimated as point values, but drawn from a Probability Distribution
    Function (PDF) whose defining metrics (like e.g. mean and standard deviation
    in a Gaussian distribution) are estimated via the model fitting process
    (see for instance \cite{jospin2022hands}, illustrated in Figure \ref{fig::pnn}).
    In this case, each time a prediction is performed, a new set of parameters
    are drawn from the PDF thus resulting in different outputs ($y$) for the same
    input ($x$) and generating an ensemble of estimates from which uncertainty can be derived.

    \begin{figure}[h!]
    \centering
    \includegraphics[width=0.7\columnwidth]{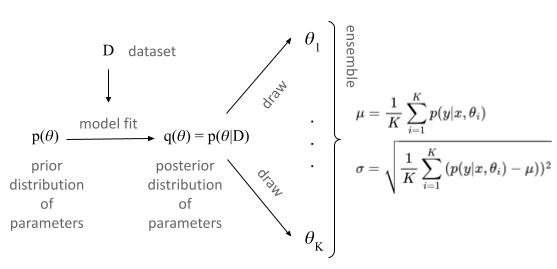}
    \caption{Probabilistic Neural Networks create various models drawing parameter values from estimated posterior distributions}\label{fig::pnn}
    \end{figure}
\end{itemize}

In addition, an important aspect to be kept in mind is that the input data used
to train the model also has certain errors (like e.g. the Vertical Total Electron Content
Root Mean Square error published in IONEX maps, see \cite{schaer1998ionex},
as used in this work). Therefore, these noise will impact the training stage and
thus the errors may need to be propagated \cite{wright1999bayesian}.

Finally, an important issue to be addressed is uncertainty calibration. The model
uncertainty is usually underestimated (when compared with actual errors) and
therefore needs to be compensated, as shown by \cite{guo2017calibration}. This
can be done using the comparison between the predicted uncertainty and the actual
error given by the e.g. validation data set. The relationship between the two can
establish a calibration function that can be subsequently used during the prediction stage.

\subsection{Ionospheric estimation using Artificial Intelligence}

Traditionally, ionospheric models have been classified into climatological models,
such as IRI (\cite{bilitza2011international}) or NeQuick (\cite{nava2008new}),
that use physical models to provide values of e.g. electron density or TEC, or
data driven models such as GNSS-based tomographic models (see for instance
\cite{roma2015real}) that use data (e.g. GNSS measurements such as pseudorange
and carrier-phase measurements) to fit a physical model of the ionosphere.
Recent advancements in neural networks have also impacted the field of ionospheric
modelling and estimation so that data can be used to fit a neural network, as proposed
already back in 1998 by \cite{cander1998artificial} and done in works such as
\cite{orus2019using} or \cite{cesaroni2020neural}. In \cite{orus2019using}, the
VTEC values from Global Ionospheric Maps in IONEX format were used as truth ($y$)
to fit a neural network whose features/inputs ($x$) were solar activity parameters,
time and location. Results show that accuracies of units of TECU (in solar minimum conditions)
could be achieved, leading to viable corrections for single-frequency users,
with similar performances than the GPS and Galileo broadcast ionospheric models.
Similar works also propose the estimation of VTEC for a wider range of ionospheric
conditions and considering Recurrent Neural Networks (RNN) that incorporate the
time correlation within the model prediction (see \cite{chen2022prediction} or \cite{xiong2021long}),
reaching accuracies in the prediction of around 4 to 5 TECU. In close connection
with Machine Learning is the continuous ingestion of massive data sets in a MLOps
(Machine Learning Operational system) context to provide with nowcast and forecast
estimates of the VTEC, as done in \cite{klopotek2024cloud}.

However, few steps have been done in order to estimate the uncertainties of the
ionospheric estimates given by a neural network with the exception of \cite{natras2023uncertainty}.
This work proposes various techniques for uncertainty estimation: one based on
what it is called a Super Ensemble, which is in fact a combination of the
“Bootstrap-aggregation” (shown above) and a selection of various neural network models
(XGBoost, SciKit-learn's Adaboost and Random forests), the other based on a modified
output layer in a Bayesian (Probabilistic) Neural Network with a two-neuron
output layer that represents the mean and standard deviation of the PDF
corresponding to the final estimate. In this latter approach, the mean is in
fact the model estimate while the standard deviation corresponds to the uncertainty of the estimate.

\section{Methodology}

This work uses the TensorFlow framework for the design and development of the
ionospheric PNN model. In order to define the model topology, there are several
hyperparameters/design choices that have been considered. These hyperparameters
are the following:

\begin{itemize}
\item Number of hidden layers: This drives the complexity of the neural network.
A number between 1 and 5 (input and output layers excluded) have been used as a starting baseline for the design.
\item Number of neuron per layers: Variable neurons have been considered,
some initial design choices were based on Table 1 of \cite{orus2019using} and Figure 2 of
\cite{natras2023uncertainty} as this allows direct comparison with results
of non-Probabilistic Neural Networks or other state-of-the-art techniques.
\item Type of layer: Within the selection of hidden layers, some were set as \emph{variational}
(i.e. whose parameters will be drawn from a density function that will be
estimated during the training stage) or not (\emph{deterministic}, i.e. parameters estimated and
fixed at training stage). This determines if the network is a purely Bayesian
(Probabilistic) Neural Network (BNN or PNN), whose layers are all probabilistic,
or a Hybrid Bayesian Neural Network (HBNN), which has a mixture of variational
and non-variational (\emph{deterministic}) layers. In this work, HBNN has been adopted,
with a balance between Probabilistic and non-Probabilistic layers.
\item Parameter distribution: The distribution of the network parameters is
in principle unknown, and multiple PDFs could be used to model them. Also,
libraries such as Tensorflow allow for various types of distributions and
even the possibility to model the cross-correlation of the different
hyperparameters using covariance matrix estimation in distributions such as
Tensorflow's \texttt{MultivariateNormalTriL} distribution. For this work, Gaussian
distributions have been adopted as a starting point for the prior and
posterior distribution of the model parameters. However, hyperparameter cross-covariance
estimation could be also considered.
\item Activation function: Rectified Linear Unit (“relu”) and “linear” activation
functions have been used for the hidden and output layer respectively.
\item Prior and posterior functions: For this work, Gaussian trainable prior
and posterior distributions have been used.
\item Batch size: Number of samples to be used in each training step. A large
batch size will speed up the training process and reduce convergence time
(less noisy steps), but could potentially lead to overfitting or getting
stuck in a non optimal minimum. A small batch size can lead to better
results in the gradient descent stage, but it will lead to a slower
training stage. A trade-off will be usually required, as shown later in the
discussion of the batch size. For this work, a batch size of 128 offers a
good compromise between training speed and performance.
\item Number of epochs: The training stage may be repeated certain number of
epochs using the same training dataset to refine and improve the convergence
step of the training stage and improve the final network estimation.
In this work, between 5 and 10 have been used for the training stage.
\end{itemize}

As a preliminary activity for the design, some BNN models have been exercised
in order to obtain a set of guidelines for the design of the network
architecture. This activity has been done for the specific use case of
ionospheric VTEC estimation. To this end, a model to test the VTEC (and associated
uncertainty) has been built and trained with IONEX data for the year 2009.
All the data for this year has been used for training, except all maps for
January 1st 2009, that have been reserved to validate the data and compare the
VTEC differences (i.e. $\varepsilon_{true} =VTEC_{BNN} - VTEC_{IONEX}$) against
the uncertainty computed by the BNN ($\sigma_{BNN}$).

The following features (inputs of the neural network) have been used to train the neural network:

\begin{itemize}
    \item Adjusted F10.7 solar flux
    \item Kp index
    \item Day of year
    \item Second of the day
    \item Longitude and latitude
\end{itemize}

Several architectures have been tested using the Tensorflow library for this
work, but the two discussed in this proposal are:
\begin{itemize}
    \item \texttt{V64-V32-V16-V1}, full BNN network, consisting of 3 hidden layers
    and 1 output layer (input layer omitted in the notation) where all layers
    are probabilistic (noted as \texttt{V} from \emph{Variational})
    \item \texttt{V64-D32-D16-D1}, hybrid BNN (HBNN), consisting of 3 hidden layers
    and 1 output layer (input layer omitted in the notation) where the first layer
    is probabilistic and the rest set to non-probabilistic (noted as \texttt{D} from \emph{Dense})
\end{itemize}

In both networks, the variational layers assume a trainable prior (p($\theta$) of Figure \ref{fig::pnn})
Gaussian distribution to minimise mismodeling due to an invalid assumption
regarding the distribution parameters of the PDF for the prior.

\section{Results and discussion}\label{sec::results}

In order to compute the VTEC and associated VTEC uncertainty, the prediction
stage of the networks under examination have been run 100 times for January 1st 2009
(i.e. the validation day). Each prediction stage involved the prediction of 62196 VTEC
values (i.e. 12 maps x 73 longitudes x 71 latitudes). The final VTEC estimates
are computed as the mean of all predictions (i.e. “ensembles”) while the uncertainty
corresponds to the standard deviation of the predictions.

\begin{figure}[h!]
\centering
\includegraphics[width=0.7\columnwidth]{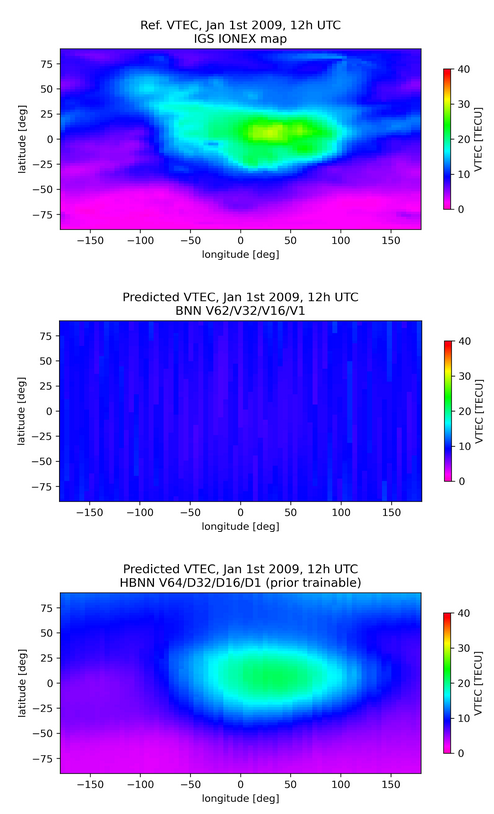}
\caption{Examples of performance using different network architectures, with
 different distributions of probabilistic and non-probabilistic layers. Top panel
 shows the reference VTEC for comparison, extracted from a single-layer IONEX map
 computed by IGS. Middle shows a full Bayesian Neural Network (all layers probabilistic,
 based on architecture \texttt{V64-V32-V16-V1}).  Bottom shows a hybrid BNN (HBNN),
 based on architecture \texttt{V64-D32-D16-D1} where the first layer is probabilistic
 (i.e. “V”)}\label{fig::results_net_arch}
\end{figure}

\begin{figure}[h!]
\centering
\includegraphics[width=0.7\columnwidth]{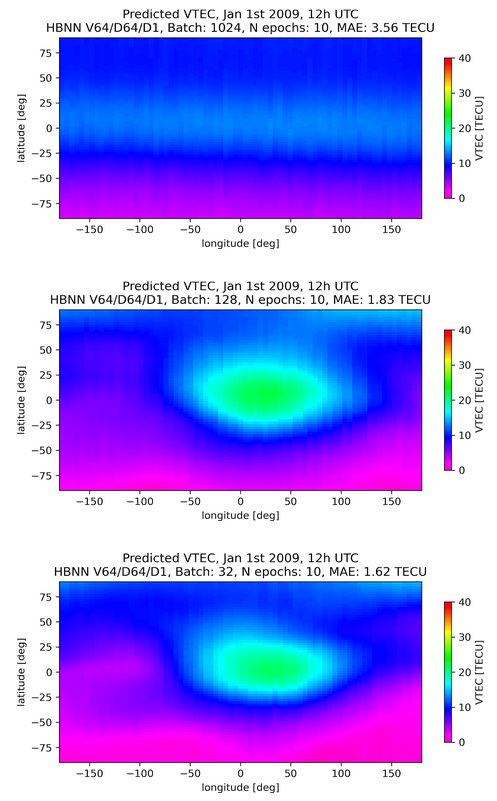}
\caption{Impact of the batch size in the training stage in a HBNN with 2
hidden layers (the first one being probabilistic). A large batch size
(e.g. 1024, top panel) might lead to incorrect results, while reducing them
too much will show very similar results (middle and bottom panels,
for batch sizes 128 and 32)}\label{fig::results_batch_size}
\end{figure}

The first architecture has been used to illustrate the fact that a HBNN might
 be the choice (rather than a pure BNN network) in the case of the ionospheric
 VTEC estimation (see Figure \ref{fig::results_net_arch}), while the second architecture is used to
 illustrate the importance of the batch size in the configuration of the training
 stage (see Figure \ref{fig::results_batch_size}).

\begin{figure}[h!]
\centering
\includegraphics[width=0.7\columnwidth]{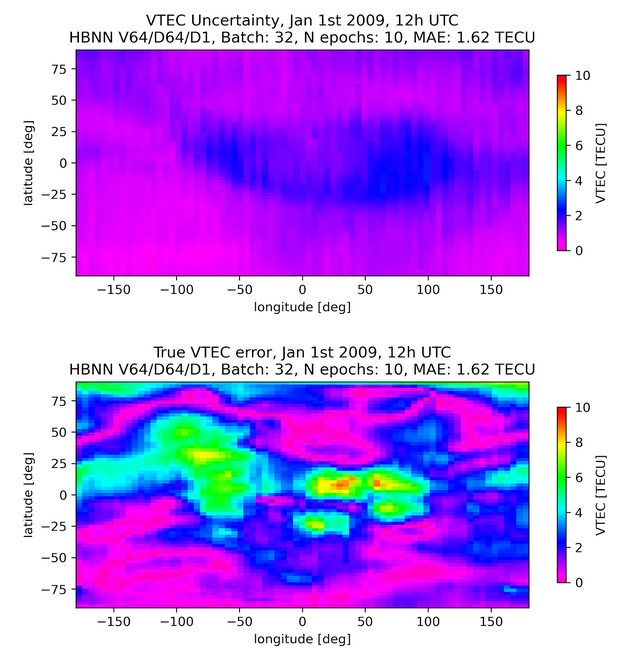}
\caption{(Top) Example of VTEC uncertainty give by one of the HBNN networks under
examination (details on the subtitle) and (bottom) actual VTEC difference (error)
as compared with the reference VTEC from IONEX map.}\label{fig::results_error_map}
\end{figure}

A key advantage of BNNs is their capacity to quantify uncertainty. Figure \ref{fig::results_error_map}
includes an example of uncertainty estimation for a single map of January 1st
2009 (12h UTC) and how it compares with the actual error (obtained as the absolute
value of the difference between the VTEC from the BNN and the VTEC for the corresponding
IONEX map).

\begin{figure}[h!]
\centering
\includegraphics[width=0.7\columnwidth]{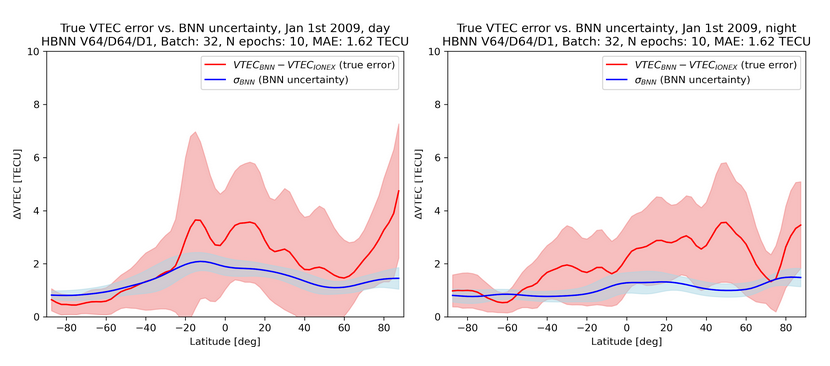}
\caption{Comparison between the true VTEC error (red) and the BNN uncertainty
(blue) vs latitude for (left) day and (right) night periods. Note: MAE stands
for Maximum Absolute Error}\label{fig::results_error}
\end{figure}

For a more quantitative plot, the latitudinal dependency of this error (discriminated
between day and night periods, considering all maps) is shown in Figure \ref{fig::results_error},
where the uncertainty and error ranges are also shown. The plot indicates that
the uncertainty approximates reasonably well the actual error, albeit a certain
calibration step is required. This has been already pointed out in several
works (see for instance \cite{natras2023uncertainty} and \cite{guo2017calibration}).

\section{Conclusion}\label{sec::conclusions}

This work has shown a preliminary application of Probabilistic Neural Networks
(PNNs) for the estimation of global Vertical Total Electron Content (VTEC) maps.
The PNN approach yielded VTEC estimations with uncertainties within the range of
established methodologies, typically a few TECU. However, a systematic bias was
identified, wherein the PNN's formal uncertainty significantly underestimated
the actual estimation error. This underestimation exhibited latitude dependence,
with the largest discrepancies observed in lower latitude regions, reaching up
to a factor of two. Therefore, future research should prioritize the calibration
of the PNN model to accurately represent the uncertainty associated with its VTEC
estimations.

\section{Usage of artificial intelligence}\label{sec::ai}

Artificial intelligence (AI) tools were used to assist with stylistic polishing and
grammar correction in some sections of this paper (using the draft written by the author).
The author subsequently reviewed and edited all AI-generated text. The core
research, including ideation, code implementation, and analysis, was conducted
entirely by the author.

\bibliography{bibliography}

\begin{thebibliography}{23}
\providecommand{\natexlab}[1]{#1}
\providecommand{\url}[1]{{#1}}
\providecommand{\urlprefix}{URL }
\providecommand{\doi}[1]{\url{https://doi.org/#1}}
\providecommand{\eprint}[2][]{\url{#2}}
 \bibcommenthead

\bibitem[{Bilitza et~al(2011)Bilitza, McKinnell, Reinisch, and
  Fuller-Rowell}]{bilitza2011international}
Bilitza D, McKinnell LA, Reinisch B, et~al (2011) The international reference
  ionosphere today and in the future. Journal of Geodesy 85(12):909--920

\bibitem[{Braiek and Khomh(2019)}]{braiek2019tfcheck}
Braiek HB, Khomh F (2019) Tfcheck: A tensorflow library for detecting training
  issues in neural network programs. In: 2019 IEEE 19th international
  conference on software quality, reliability and security (QRS), IEEE, pp
  426--433

\bibitem[{Cander(1998)}]{cander1998artificial}
Cander LR (1998) Artificial neural network applications in ionospheric studies.
  ANNALI DI GEOFISICA 5(6)

\bibitem[{Cesaroni et~al(2020)Cesaroni, Spogli, Aragon-Angel, Fiocca, Dear,
  De~Franceschi, and Romano}]{cesaroni2020neural}
Cesaroni C, Spogli L, Aragon-Angel A, et~al (2020) Neural network based model
  for global total electron content forecasting. Journal of space weather and
  space climate 10:11

\bibitem[{Chen et~al(2022)Chen, Liao, Li, Wang, Deng, and
  Hong}]{chen2022prediction}
Chen Z, Liao W, Li H, et~al (2022) Prediction of global ionospheric tec based
  on deep learning. Space Weather 20(4):e2021SW002854

\bibitem[{Gawlikowski et~al(2023)Gawlikowski, Tassi, Ali, Lee, Humt, Feng, and
  Kruspe}]{gawlikowski2023survey}
Gawlikowski J, Tassi CRN, Ali M, et~al (2023) A survey of uncertainty in deep
  neural networks. Artificial Intelligence Review 56(Suppl 1):1513--1589

\bibitem[{Guo et~al(2017)Guo, Pleiss, Sun, and Weinberger}]{guo2017calibration}
Guo C, Pleiss G, Sun Y, et~al (2017) On calibration of modern neural networks.
  In: International conference on machine learning, PMLR, pp 1321--1330

\bibitem[{Hüllermeier and Waegeman(2021)}]{hullermeier2021aleatoric}
Hüllermeier E, Waegeman W (2021) Aleatoric and epistemic uncertainty in
  machine learning: An introduction to concepts and methods. Machine learning
  110(3):457--506

\bibitem[{Jospin et~al(2022)Jospin, Laga, Boussaid, Buntine, and
  Bennamoun}]{jospin2022hands}
Jospin LV, Laga H, Boussaid F, et~al (2022) Hands-on bayesian neural
  networks—a tutorial for deep learning users. IEEE Computational
  Intelligence Magazine 17(2):29--48

\bibitem[{Juan et~al(2025)Juan, Timote, Sanz, Rovira-Garcia,
  Gonz{\'a}lez-Casado, Aragon-Angel, Yin, Garc{\'\i}a-Mateos,
  Or{\'u}s-P{\'e}rez, and Fern{\'a}ndez-Hern{\'a}ndez}]{juan2025iono4has}
Juan JM, Timote CC, Sanz J, et~al (2025) Iono4has, a real-time ionospheric
  model for galileo high accuracy service, sl2. results and validation. In:
  Proceedings of the 2025 International Technical Meeting of The Institute of
  Navigation, pp 805--818

\bibitem[{Kłopotek et~al(2024)Kłopotek, Pan, Sturn, Weinacker, See, Crocetti,
  and Awadaljeed}]{klopotek2024cloud}
Kłopotek G, Pan Y, Sturn T, et~al (2024) A cloud-native approach for
  processing of crowdsourced gnss observations and machine learning at scale: A
  case study from the camaliot project. Advances in Space Research

\bibitem[{Lee et~al(2020)Lee, Ullah, and Wang}]{lee2020bootstrap}
Lee TH, Ullah A, Wang R (2020) Bootstrap aggregating and random forest. In:
  Macroeconomic forecasting in the era of big data: Theory and practice. p
  389--429

\bibitem[{Mohebali et~al(2020)Mohebali, Tahmassebi, Meyer-Baese, and
  Gandomi}]{mohebali2020probabilistic}
Mohebali B, Tahmassebi A, Meyer-Baese A, et~al (2020) Probabilistic neural
  networks: a brief overview of theory, implementation, and application. In:
  Handbook of probabilistic models. p 347--367

\bibitem[{Natras et~al(2023)Natras, Soja, and Schmidt}]{natras2023uncertainty}
Natras R, Soja B, Schmidt M (2023) Uncertainty quantification for machine
  learning‐based ionosphere and space weather forecasting: Ensemble, bayesian
  neural network, and quantile gradient boosting. Space Weather
  21(10):e2023SW003483

\bibitem[{Nava et~al(2008)Nava, Coisson, and Radicella}]{nava2008new}
Nava B, Coisson P, Radicella SM (2008) A new version of the nequick ionosphere
  electron density model. Journal of atmospheric and solar-terrestrial physics
  70(15):1856--1862

\bibitem[{Orús-Perez(2019)}]{orus2019using}
Orús-Perez R (2019) Using tensorflow-based neural network to estimate gnss
  single frequency ionospheric delay (iononet). Advances in Space Research
  63(5):1607--1618

\bibitem[{Roma-Dollase et~al(2015)Roma-Dollase, López~Cama,
  Hernández~Pajares, and García~Rigo}]{roma2015real}
Roma-Dollase D, López~Cama JM, Hernández~Pajares M, et~al (2015) Real-time
  global ionospheric modelling from gnss data with rt-tomion model. In: 5th
  International Colloquium Scientific and Fundamental Aspects of the Galileo
  Programme, Braunschweig, Germany, pp 27--29

\bibitem[{Schaer et~al(1998)Schaer, Gurtner, and Feltens}]{schaer1998ionex}
Schaer S, Gurtner W, Feltens J (1998) Ionex: The ionosphere map exchange format
  version 1. In: Proceedings of the IGS AC workshop, Darmstadt, Germany

\bibitem[{Specht(1990)}]{specht1990probabilistic}
Specht DF (1990) Probabilistic neural networks. Neural networks 3(1):109--118

\bibitem[{Ståhl et~al(2020)Ståhl, Falkman, Karlsson, and
  Mathiason}]{stahl2020evaluation}
Ståhl N, Falkman G, Karlsson A, et~al (2020) Evaluation of uncertainty
  quantification in deep learning. In: International Conference on Information
  Processing and Management of Uncertainty in Knowledge-Based Systems, Springer
  International Publishing, pp 556--568

\bibitem[{Wright(1999)}]{wright1999bayesian}
Wright WA (1999) Bayesian approach to neural-network modeling with input
  uncertainty. IEEE Transactions on Neural Networks 10(6):1261--1270

\bibitem[{Xiong et~al(2021)Xiong, Zhai, Long, Zhou, Zhang, and
  Shen}]{xiong2021long}
Xiong P, Zhai D, Long C, et~al (2021) Long short‐term memory neural network
  for ionospheric total electron content forecasting over china. Space Weather
  19(4):e2020SW002706

\bibitem[{Zumberge et~al(1997)Zumberge, Heflin, Jefferson, Watkins, and
  Webb}]{zumberge1997precise}
Zumberge JF, Heflin MB, Jefferson DC, et~al (1997) Precise point positioning
  for the efficient and robust analysis of gps data from large networks.
  Journal of geophysical research: solid earth 102(B3):5005--5017

\end{thebibliography}

\end{document}